\documentstyle[12pt,aasms4]{article}

\newcommand\cA{{\cal A}}
\newcommand\mt{{\cal M}_t}
\newcommand\bk{{\bf k_h}}
\newcommand\bkx{{\bf k_h\cdot x}}
\newcommand\bx{{\bf x}}
\newcommand\be{\begin{equation}}
\newcommand\ee{\end{equation}}
\font\gkvec=cmmib10                         
\newcommand\bxi{\hbox{{\gkvec\char24}}}

\input epsf
\lefthead{P. Kumar and S. Basu}
\righthead{Line asymmetry of solar p-modes}
\begin{document}

\title{Line asymmetry of solar p-modes: \\
Reversal of asymmetry in intensity power spectra}

\author{Pawan Kumar and Sarbani Basu}
\affil{Institute for Advanced Study, Olden Lane, Princeton, NJ 08540, U. S. A.}

\begin{abstract}

The sense of line asymmetry of solar p-modes in the intensity power spectra 
is observed to be opposite of that seen in the velocity power spectra.
Theoretical calculations provide a good understanding and fit to the
observed velocity power spectra whereas the reverse sense of asymmetry
in the intensity power spectrum has been poorly understood. We show that 
when turbulent eddies arrive at the top of the convection zone they 
give rise to an observable intensity fluctuation which is correlated 
with the oscillation they generate, thereby affecting the shape of the 
line in the p-mode power spectra and reversing the sense of asymmetry
(this point was recognized by Nigam et al. and Roxburgh \& Vorontsov).
The addition of the correlated noise displaces the frequencies of
peaks in the power spectrum. Depending on the amplitude of the noise
source the shift in the position of the peak can be substantially
larger than the frequency shift in the velocity power spectra. In neither
case are the peak frequencies precisely equal to the eigenfrequencies of 
p-modes. We suggest two observations which can provide a test
of the model discussed here.
\end{abstract}


\section{Introduction}

Duvall et al. (1993) found that solar p-mode line profiles
are not exactly Lorentzian in shape but are somewhat asymmetric about the
peak of the power. They showed that the power in their observed
surface velocity spectra fell off more slowly at the low frequency side of 
the peak compared to the high frequency side of the peak. They also claimed
that the sense of asymmetry is opposite in the intensity power spectra
i.e. there is more power at the high frequency side of the peak compared
to the low frequency side. The results of Duvall et al. have been 
recently confirmed by a number of independent observations. 

Theoretical calculations have little difficulty in reproducing the
observed velocity power spectrum. According to the simplest models
one expects the shape of velocity and intensity power spectra to be 
almost identical, as long as both of these are observed at the same 
place in the solar atmosphere, since the velocity and temperature 
eigenfunctions of a mode are linearly related to each other. 
Rast \& Bogdan (1998) point out that
nonadiabatic effects are unlikely to modify this relation. Nor does it
appear that the process of the formation of absorption lines in the
optical spectrum could cause the asymmetry of intensity power spectrum
to be reversed (Sasselov, personal communication). It is therefore puzzling 
that the intensity power spectra are observed to be very different from the 
velocity spectra. Nigam et al. (1998) suggested that correlated noise 
added to p-mode oscillation amplitude affects the shape of the 
observed spectrum and could be responsible for the difference between 
velocity and intensity power spectra; they suggested that
granules could provide the correlation. Roxburgh \& Vorontsov (1997) suggested
that velocity variation associated with acoustic sources, i.e., the turbulent
eddies, are observed when eddies arrive at the top of the convection 
zone and this variation should be correlated with p-mode oscillations. 
Roxburgh \& Vorontsov chose, arbitrarily, the relative amplitude 
of the velocity variation associated with oscillation and with the noise 
source to be of order unity and considered them to be almost perfectly
correlated in order to produce the reversal of line asymmetry. 

We calculate the relative contribution of turbulent eddies to the observed flux
variation and its correlation with p-mode oscillation in a self consistent 
manner and show that this process explains the observed reversal of 
asymmetry in the intensity power spectra. The calculations are 
presented in section 2. The conclusion and some predictions are 
contained in section 3.

\section{Intensity oscillation and line asymmetry}

We first consider a simple one dimensional square well potential model,
same as that considered in Abrams \& Kumar 1996, to calculate the power 
spectrum of p-mode oscillations and include the contribution of 
fluctuating sources, which excite these oscillations, to the observed 
flux variation amplitude:

\be
 {d^2\psi_\omega\over d r^2} + i2\omega\gamma\psi_\omega + 
   \left[\omega^2 - V(r)\right]\psi_\omega = 
   S_\omega\, \delta(r-r_s), 
\ee
where $\psi_\omega$ is the temperature perturbation function, $S_\omega$ 
is the quadrupole source amplitude, $\gamma$ is the damping constant which
is taken to be independent of $r$, and $V(r)$ is a simplified effective 
potential constructed to describe the solar p-modes: $V=\infty$ for
$r\le 0$, for $0<r<a$ the potential is zero, and $V(r)=\alpha^2$, a constant,
for $r \ge a$; $a$ is the sound travel time from the lower to the upper 
turning point of a given mode and $\alpha$ is the acoustic cutoff frequency 
at the temperature minimum. The observed flux variation is roughly proportional 
to the Lagrangian temperature perturbation at the place in the atmosphere
where lines are formed\footnote{A perturbation of the convective flux by 
oscillations could however modify the relation between the observed flux
variation and the Lagrangian temperature perturbation.}. 

For sources lying in the evanescent region i.e. $r_s\ge a$, the amplitude 
of the temperature variation observed at point $r_o$ can be easily
calculated by solving Eq.~(1) and is given by

\be
 T_\omega(r_o) = S_\omega G_\omega(r_s,r_o) + {S_\omega\beta\over\omega_2}
 \exp(i\phi-\eta),
\ee
where
\be
 G_\omega(r_s,r_o) = - {\exp\bigl[-\omega_2(r-a)\bigr]\over\omega_2}
   \left[ {\omega_2\sin(\omega_1 a) \cosh(\omega_2\delta r) + \omega_1
   \cos(\omega_1 a)\sinh(\omega_2\delta r)\over \omega_2\sin\omega_1 a +
   \omega_1\cos\omega_1 a} \right], 
\ee
is the Green's function for Eq.~(1), 
$\omega_1 = (\omega^2 + i2\gamma\omega)^{1/2}$, $\omega_2 = (\alpha^2 - 
\omega^2 - i2\gamma\omega)^{1/2}$, $\delta r = r_s - a$, 
and $\eta$ is the optical depth between $r_o$,
the point where the observation is made, and the top of the convection zone.
The dimensionless parameter $\beta$ is the ratio of the observed amplitude 
of the flux variation of an eddy  to that  of the oscillation, and $\phi$ 
is its phase difference with respect to the mode it excites at radius $r_o$.

The real and the imaginary parts of the Green's function have very
different shapes as shown in fig. 1, and this enables a correlated
noise source to modify the observed shape of the power spectrum as 
discussed below. 

In the absence of the contribution of the noise source to the observed
flux variation, i.e. when $\beta=0$, the shape of the power spectrum is 
shown in the left panel of Fig. 2. This case should
correspond to the velocity power spectrum of solar p-modes since the random
component of the velocity field in the solar atmosphere ---  where optical 
observations are made --- is believed to be uncorrelated with the turbulent 
eddies in the convection zone which excite p-mode oscillations. The asymmetry
in our calculated spectrum does in fact look very much like that seen 
in the velocity power spectra of solar p-modes.

Unlike the velocity variations,
the flux variation associated with turbulent eddies are observed
when the eddies arrive at the top of the convection zone. These variations
are correlated with oscillations these eddies generate; this correlation
is parameterized above by $\beta$ and $\phi$. The addition of this `noise'
term to the oscillatory signal modifies the relative contribution of the
real and the imaginary parts of the Green's function to the observed
intensity spectrum (which is proportional to $|T_\omega(r_o)|^2$) and 
thereby modifies the shape of the p-mode lines in the intensity
observations (see Fig.~2, right panel). We find the reversal of asymmetry over a 
wide range of $\beta$ and $\phi$ as shown in fig. 3; these line profiles
are similar to the observed reversal of asymmetry seen
in the solar intensity data. Note that the background in the intensity  
spectrum is significantly non-zero, and in fact if this model is correct
we expect the reversal of asymmetry to disappear when the background
in the spectrum vanishes.

The addition of the correlated noise source to the observed flux
variation significantly modifies frequencies of peaks in the power
spectrum as is clearly seen in Fig.~3. A change in the amplitude of
the noise source, as quantified by the dimensionless parameter $\beta$, 
from 0.5 to 2 shifts the peak frequency of 2.4mHz mode by about 
0.4 $\mu$Hz. The peak frequencies in the intensity power spectra 
are different from the peaks in the velocity spectra and neither 
of these are the true eigenfrequencies of p-modes.
The frequency shift in the intensity spectra depends on the amplitude
of the background in addition to the effects that cause shift in the velocity 
spectra (cf. Abrams \& Kumar 1996, Christensen-Dalsgaard et al. 1998), and so 
it is safer to use the velocity spectra to determine the true eigenfrequencies 
of solar p-modes. 

We calculate the parameters $\beta$ and $\phi$ for a realistic model below.

\subsection{ Relation between wave and source amplitudes}

We consider a plane parallel atmosphere which sits in a constant gravitational
field, ${\bf g}$, and consists of two semi-infinite
layers, the lower layer being adiabatic and polytropic, and the upper
isothermal. The pressure, $p$, density, $\rho$, and temperature,
$T$, are continuous across the interface between the two layers.
In the lower layer the adiabatic and polytropic indices are
related by $\gamma=1+1/m$. The $z$ coordinate
measures {\it depth} below the level at which the adiabatic layer
would terminate in the absence of the isothermal layer.
In the adiabatic layer the thermodynamic variables have a power law 
dependence on the depth $z$ e.g. $p=p_t (z/z_t)^{m+1}$, $\rho=\rho_t (z/z_t)^m$,
the sound speed $c^2 = gz/m$, and the scale height $H=z/(m+1)$, where
the subscript $t$ denotes quantities evaluated at the top of the adiabatic
layer.
In the isothermal atmosphere $p$ and $\rho$ are proportional to $\exp(z/H_i)$.

Following Goldreich and Kumar (1990), hereafter referred to as GK,
we use $Q=P_1/\rho$ as the wave variable,
where $P_1$ is the Eulerian pressure perturbation. The inhomogeneous
wave equation in the adiabatic atmosphere is

\be
 {d^2 Q\over dz^2} + {g\over c^2}{dQ\over dz} +
  -{1\over c^2}{\partial^2 Q\over \partial t^2} - k_h^2 Q = 
  - {\partial^2 \over \partial t^2} \left( {s\over c_p} \right) - 
  {g\over \rho} {\partial\over\partial z} \left( {\rho s\over c_p}\right)
  \equiv S,
\ee
where $s$ is the Eulerian entropy perturbation associated with the
turbulent convection. The source term $S$ consists of a monopole and
a dipole piece which arise as a result of volume change due to
fluctuation in entropy and buoyancy force variation associated with 
this volume change respectively. For simplicity we have not included the 
Reynold's stress source term associated with fluctuating internal stresses 
in the convection zone. Inclusion of Reynold's stress
makes the algebra considerably more tedious but does not modify the
main results of this section in any essential way.\footnote{ GK 
showed that the monopole and the dipole source terms in $S$ cancel
leaving the remainder which has magnitude of same order as the quadrupole
Reynold's term.}

We expand the enthalpy wave function $Q$ in terms of the normal modes, 
$Q_\alpha(z)$, which are normalized to unit energy,

\be
 Q = {1\over \sqrt{2\cA}} \sum_\alpha {\left[A_\alpha Q_\alpha
\exp(-i\omega t + i\bkx) + A^*_\alpha Q_\alpha^* \exp(
i\omega t - i\bkx)\right]}, 
\ee
where $\cA$ is the horizontal cross section area of the
atmosphere, and $\bk$ is the horizontal wave vector.
Substituting this expansion into equation (4) and multiplying both
sides with $Q^*_\alpha\exp(i\omega t - i\bkx)$, and integrating
over space and time, we obtain

\be
A_\alpha(t) = {1\over 2i\omega \cA^{1/2}} \int_{-\infty}^t
dt\int d^3x\, Q^*_\alpha\, S \exp(i\omega t - i\bkx). 
\ee
Using the wave equation (4) this can be rewritten as

\be
 A_\alpha(t) \approx -{1\over 2i\omega \cA^{1/2}}
  \int^t_{-\infty} dt\int d^3x\, {\rho c^2 s\over c_p}\, {\partial^2 
  Q^*_\alpha \over\partial z^2} \exp(i\omega t - i\bkx). 
\ee

The change to mode amplitude, $\delta A_\alpha$, due to a single eddy 
of size $h$ over its lifetime $\tau_h$, which is located at $\bf x_0$
and lived at time $t_0$, can be easily estimated from the above equation 
and is given by

\be
 \delta A_\alpha(t) \approx -{\rho c^2 h^3 \tau_h 
  s_h(t_0) \over 2i\omega 
  c_p \cA^{1/2}} {\partial^2 Q^*_\alpha\over \partial z^2}\,\exp\bigl[
  i\omega_\alpha t_0 - i{\bf k_h\cdot x_0} -\gamma_\alpha(t-t_0)
  + i\phi_1\bigr],
\ee
where the phase $\phi_1$ depends on the temporal and spacial
properties of the eddy, and the factor $\exp[-\gamma_\alpha(t-t_0)]$ 
has been multiplied to model wave damping that causes the change to mode
amplitude to decrease with time for $t\ge t_0$.
GK showed that the resonant scale height size eddies are most efficient 
at exciting low frequency modes (modes of frequency less than about 3mHz); 
for these eddies $h\sim H$ and $\tau_h\sim 
\omega_\alpha^{-1}$. The depth ($z_\omega$) of these eddies below the
photosphere, or the isothermal atmosphere in our model, can be 
estimated from the constancy of the convective flux $\rho v^3$

\be
 z_\omega \sim z_t \left[{\mt\omega_{ac}\over\omega}\right]^{3/(m+3)}, 
\ee
where $\mt$ is the Mach number of the turbulent convection at the
top of the convection zone, and $\omega_{ac}$ is the acoustic cut
off frequency in the isothermal atmosphere.

The change to the enthalpy perturbation associated with a mode $\alpha$, 
$\delta Q_\alpha(z_{o})$, in the isothermal atmosphere at location $z_{o}$ 
due to the eddy is given by

\be
 \delta Q_\alpha(z_{o},t) = {\delta A_\alpha Q_\alpha(z_{o})
  \exp(-i\omega_\alpha t + i \bkx)\over (2\cA)^{1/2}}. 
\ee
Substituting for $\delta A_\alpha$ from Eq.~(8) we find

\be
\delta Q_\alpha(z_{o},t) = i{\rho(z_\omega)c^2(z_\omega) H_\omega^3\over 
   2^{3/2}\cA\omega^2} {\delta T_{ed}(z_\omega,t_0)\over T(z_\omega)} 
   Q_\alpha(z_{o}) {\partial^2 Q^*_\alpha\over \partial z^2}
   \exp\Bigl[ -(\gamma_\alpha+i\omega_\alpha)(t-t_0) + i\phi_1 + 
  i{\bf k_h\cdot(x-x_0)}\Bigr],
\ee
where $\delta T_{ed}=s_h/c_p$ is the temperature fluctuation of the eddy.
Low frequency p-modes are most likely excited near the top of the 
convection zone in a region where waves are evanescent (cf. GK).
The wave function is nearly constant in the evanescent region and its
derivative $\partial^2 Q_\alpha/\partial z^2 \sim (\omega^4/g^2) Q_\alpha$.
Moreover, the normalized eigenfunction in the evanescent region is
(see GK)

\be
 Q_\alpha \sim {z_t^{m/2} \omega^{(m-1)} k_h^{1/2} \over \rho_t^{1/2} 
   g^{(m-2)/2} }. 
\ee

Making use of these equations, the expression for $\delta Q_\alpha$ reduces to

\be
 \delta Q_\alpha(z_{o},t) \sim {i k_h z_\omega^{m+3}\omega^{2m}\over\cA
    g^m} c_p\delta T_{ed}(z_\omega, t_0) \exp\bigl[ -(\gamma_\alpha +
   i\omega_\alpha)(t-t_0) + i {\bf k_h\cdot(x-x_0) + i\phi_1} \bigr].
\ee
Finally substituting for $z_\omega$ from Eq.~(9) and using
the relation $g/z_t\approx \omega^2_{ac}$, the above equation further
simplifies to

\be
 \delta Q_\alpha(z_{o},t) \sim {i k_h z_t^3\over\cA} \mt^3
   \left( {\omega\over\omega_{ac}}\right)^{2m-3} c_p\delta 
   T_{ed}(z_\omega, t_0) \exp\bigl[ -(\gamma_\alpha+i\omega_\alpha)(t-t_0) 
  + i \bk\cdot(\bx-\bx_0) + i\phi_1\bigr]. 
\ee

We next consider the relation between $\delta Q$ and the associated change
in the temperature amplitude of the mode at the height in the atmosphere 
where optical observations are made. For the nearly isothermal solar 
atmosphere the Lagrangian temperature perturbation and the pressure 
perturbations are related by

\be
 \Delta T = \Delta p \left({\partial T\over\partial p}\right)_s \approx
   {\Delta p\over\rho c_p} = {p_1 + \rho g \xi_z\over\rho c_p},
\ee
where $\xi_z$ is the radial displacement associated with the wave, and
for low frequency acoustic waves it is given by (see GK for detail)

\be
 \xi_z \approx -{1\over\omega_{ac}^2}\left[ {dQ\over dz} + {(\gamma-1)Q\over
   \gamma H_i}\right] \approx -{Q\over H_i \omega_{ac}^2}\left[ 
   {\gamma-1\over\gamma} - {\omega^2\over 4\omega_{ac}^2}\right]. 
\ee
Therefore,

\be
 \Delta T \approx {\omega^2\over\omega_{ac}^2} {Q\over c_p}.
\ee

Substituting this into Eq.~(14) we obtain the change to the Lagrangian 
temperature perturbation in the isothermal atmosphere as a result 
of wave generated by the entropy fluctuation of an eddy

\be
 \Delta T_\alpha(z_{o},{\bf x}_h,t)\approx i F(\omega, k_h)
   \delta T_{ed}(z_\omega, t_0)\exp\bigl[ -(\gamma_\alpha+i\omega_\alpha)
   (t-t_0) + i {\bf k_h\cdot(x-x_0)} + i\phi_1 \bigr],
\ee
where $F(\omega, k_h)$ is a dimensionless function defined by

\be
 F(\omega, k_h) = {k_h z_t^3\over \cA}\left({\omega\over \omega_{ac}}
   \right)^{2m-1} \mt^3. 
\ee

The total observed temperature fluctuation in the atmosphere, as seen
in a line or continuum, at some location on the solar disk, consists
of the contribution from oscillations and the attenuated emission from
turbulent eddies at the top of the convection zone. 

\begin{eqnarray}
 \delta T_a(z_{o},{\bf x}_h,t)& = & i\sum_{eddies} F(\omega,k_h) 
   \delta T_{ed}(z_\omega, {\bf x}_0, t_0) \exp\bigl[ -(
   \gamma_\alpha + i\omega_\alpha)(t-t_0) 
  + i {\bf k_h\cdot(x-x_0)} + i\phi_1\bigr] \nonumber \\
  & & \quad\quad\quad + \delta T_{ed}(z_{t},{\bf x},t)\exp(-\eta), 
\end{eqnarray}
where $\delta T_{ed}(z_{t},{\bf x},t)$ is the temperature variation
associated with eddies at the top of the convection zone, and $\eta$
is the optical depth of the line, used in the observation, at the
base of the photosphere.

The summation above is carried out to include all resonant eddies.
To project out modes of a particular horizontal wave vector
we multiply both sides by $\exp(-i{\bf k_h\cdot x})$ and integrate over the
area $\cA$ of the box. Individual modes are isolated by 
Fourier transforming in time and the result is

\begin{eqnarray}
 { \delta T_a(z_{o},\bk,\omega)}& = & {-\cA\over 
   (\omega- \omega_\alpha) + i\gamma_\alpha} \sum_{eddies} 
  F(\omega,k_h) \delta T_{ed}(z_\omega,\bx_0,t_0) \exp(i\omega t_0 - 
  i\bk\cdot \bx_0 + i\phi_1)  \nonumber\\
&  & \quad\quad\quad + \tau_t z_t^2 \sum_{eddies} \delta T_{ed}(z_{t},
   {\bf x}_1, t_1) \exp(i \omega t_1 -i\bk\cdot\bx_1 + i\phi_1'-\eta),
\end{eqnarray}
where the phase $\phi_1'$ depends on the details of the 
spacial and temporal structure of eddies, and it should be
roughly equal to the phase $\phi_1$ so long as the temporal 
property of the eddy does not change dramatically between the 
depth where the waves are excited ($z_\omega$)
and the top of the convection zone.

The observed power spectrum $P(\omega)=|\delta T_a|^2$.
Assuming that $\delta T_{ed}$ for different eddies are uncorrelated, 
we retain only those cross terms in $|\delta T_a|^2$ which 
correspond to the same eddy; $(t_0-t_1)$ in this case is equal 
to $(z_\omega-z_t)/v_{ed}\equiv \delta t_{tt}$ is the 
time for downward moving eddies to travel from the top of the convection 
zone to the depth where they most efficiently excite the mode. The power 
spectrum at the frequency $\omega_\alpha$ is thus given by

\be
 P(\omega_\alpha)\approx {\cA^3\over z_t^2\gamma_\alpha^3\tau_t} \left| 
   F(\omega_\alpha,k_h)\delta T_{ed}\right|^2 - {\cA^2\over \gamma_\alpha^2} 
   F(\omega_\alpha,k_h) \sin(\omega_\alpha\delta t_{tt}) \left|
   \delta T_{ed}\right|^2 \exp(-\eta), 
\ee
In deriving the above equation we have assumed that the duration of 
the observation is greater than the mode lifetime; for smaller observational 
time $T$, the second term is reduced by a factor of $T\gamma_\alpha$.

Due to the multiplicative factor $\sin(\omega_\alpha\delta t_{tt})$ the 
second term in equation (16) averages to zero if downward and upward 
moving eddies contribute 
equally to wave excitation. Thus the reversal of asymmetry seems to 
require that downward moving eddies contribute more to the generation 
of acoustic waves; some recent observations suggest that modes are 
probably excited by the downflows (cf. Goode et al. 1998).

Comparing the two terms in Eq.~(21) with the corresponding
terms in equation (2) we find that the phase difference 
between the observed flux variation of a mode and the turbulent eddy is
$\phi = \pi - \omega\delta t_{tt}$; In deriving this result we expanded
the Green's function in equation (3)  in the neighborhood of an
eigenmode: $G_\omega \approx \omega[(\omega-\omega_\alpha) +
 i\gamma_\alpha]^{-1}/(8a\alpha^2)$.
The value of $\delta t_{tt}$ is shown in fig. 4, and we see that $\phi\sim
1$ rad for low frequency modes.

Moreover, from equations (21) and (2) we find

\be
 \beta \approx {\omega_\alpha\tau_t\over 8(a\omega_{ac})k_h z_t}
  \left({\omega_{ac}\over \omega_\alpha} \right)^{2m-1} \mt^{-3}.
\ee
At the top of the solar convection zone $\mt\sim 0.3$, $z_t\sim 300$km, 
$g\sim 2.77$x10$^4$ cm s$^{-1}$, $m\sim 1.5$, $\tau_t\sim 100$s, 
and in the solar photosphere $\omega_{ac}\sim$ 3.3x$10^{-2}$rad s$^{-1}$. 
The sound travel time from the lower to the upper turning 
point, $a$, for this model atmosphere is approximately
$\omega_\alpha/(g k_h)$, and therefore $a k_h\sim \omega_\alpha/g$.
Substituting all this into equation (23) we find the value of $\beta$ 
to be about 100 for a mode of frequency 2mHz, and this is independent
of the horizontal wave number $k_h$. A more realistic model of the sun 
considered in the next subsection yields essentially the same value 
for $\beta$ when sources are taken to lie at a depth $z_\omega\sim 100$km 
as suggested by the mixing length theory (see eq. 9). For this source
depth the line asymmetry in intensity power spectra is reversed as long as
the effective optical depth, at the top of the convection zone, for the
wavelength band used in the intensity measurement is less than about 5 so that 
$\beta'=\beta\exp(-\eta)$ is greater than $\sim 0.5$ (see fig. 3).

\subsection{ Results for a solar model}

The equation for $\beta$ for a solar model, corresponding to equation (23) of
the plane parallel atmosphere, can be derived in a manner analogous to that
given in \S2.1 and so we outline a few of the main steps in the derivation.

It is convenient to work with the displacement wave function in this
more general case, instead of the enthalpy perturbation used in \S2.1,
which we expand in terms of the complete set of eigenfunctions 
$\bxi_\alpha Y_{\ell m}$

\be 
 \bxi = {1\over \sqrt{2}} \sum_\alpha A_\alpha \bxi_\alpha Y_{\ell m}
   \exp(-i\omega_\alpha t) + c.c.
\ee
The change to the radial displacement amplitude of a mode, due to one eddy,
at radius $r_o$ in the solar atmosphere is given by (see Goldreich et al. 1993) 

\be 
\delta\xi_r(r_o) \approx {i\tau_H\omega_\alpha H_\omega^3\over 2}\left({\partial 
  p\over \partial s}\right)_\rho s_H \xi_{\alpha r}(r_o) \left( 
  {\partial\xi_{\alpha r}^* \over\partial r}\right)_{r_\omega} Y_{\ell m}^*.
\ee
The notation used here is same as in equation (11). The change to the Lagrangian 
temperature perturbation in the atmosphere resulting from the excitation due
to one eddy is calculated using the above equation and is given by

\be
\Delta T_\alpha(r_o) \approx -{i\omega^2 H_i H_\omega^3\over 2(\gamma-1) T_\omega}
    \left({\partial p\over \partial s}\right)_\rho \xi_{\alpha r}(r_o) \left(
   {\partial\xi_{\alpha r}^* \over\partial r}\right)_{r_\omega} Y_{\ell m}^*
   \delta T_{ed},
\ee
where $H_i$ is the pressure scale height at the place in the atmosphere where
the optical observation is made, $H_\omega$ is the scale length of resonant
eddies, and $r_\omega$ is the radius where most of the wave excitation takes 
place. It is straightforward to calculate $\beta$ from this equation

\be
\beta \approx {2(\gamma-1) H_t^2 c_v(r_\omega)\over H_i H_\omega^3 R_\odot^2
     \omega^2_\alpha} \left({\partial T\over \partial p}\right)_\rho
    \left[ \xi_{\alpha r}(r_o)\left({\partial\xi_{\alpha r}^* \over
    \partial r}\right)_{r_\omega}\right]^{-1} {\omega_\alpha \tau_t\over
   a\omega_{ac}},
\ee
where the subscript $t$ refers to quantity evaluated at the top of the
convection zone, and $a$ is the sound travel time from the lower to the
upper turning point of the wave. Figure 4 shows graphs for $\beta$, for 
three p-modes of a solar model, as a function of depth below the photosphere. 
The value of $\beta$ for low frequency modes, at reasonable excitation depth, 
is consistent with our estimate of \S2.1. Also shown in fig. 4 is the
eddy travel time, $\delta t_{tt}$, from the top of the convection zone,
from which we can estimate the phase difference $\phi=\pi-\omega_\alpha
\delta t_{tt}$. The values of $\beta$ and $\phi$ for the solar model seem
to lie within the parameter range needed for the reversal of line asymmetry.

Specifically, if a mode of frequency $\sim 2$ mHz is excited by quadrupole
sources lying about 100 km below the top of the convection zone then $\beta\sim 
100$ (see fig. 4); this is same as estimated in \S2.1. For flux variation
observations using the central part of the Ni line, one of the filter bands
used by the GONG and the MDI instruments, the value of $\eta$, the optical 
depth at the top of the convection zone, is about 6.
Thus $\beta'=\beta\exp(-\eta)$ for this source depth and the optical line
is about 0.3, and figs. 3 and 4 suggest that the line asymmtery in the 
intensity power spectra should be opposite to the velocity power spectra. 
However, if the acoustic sources are located at a depth
of 200 km or greater, as suggested by Kumar and Basu (1998), then $\beta$ 
would be much smaller (see fig. 4) and the line asymmetry in the intensity
power spectra obtained using the Ni line should be the same as the velocity
spectra. The line asymmtery in intensity power spectra, for deeper lying 
sources, is expected to be reversed when observations are made in the
continuum for which $\eta\sim 1$. 

The asymmetry reversal requires that the effective observed 
amplitude of the noise source, $\beta'\equiv\beta\exp(-\eta)$, 
exceed some minimum value of order 0.1 (see fig. 3). Thus 
we expect the velocity and the intensity power spectra to have 
the same sense of asymmetry, irrespective of the source depth, 
in those observations carried out using optical lines formed high 
up in the chromosphere so that their optical depth $\eta$ at the 
top of the convection zone is large. This prediction provides a 
verifiable test of the model discussed here.

For a fixed source depth the value of $\beta$ decreases rapidly 
with increasing mode frequency (see fig. 4). Thus if modes of
different frequencies were to be excited at the same depth, the
line asymmetry reversal seen in the intensity power spectra should 
disappear at some frequency of order 3.5 mHz. Asymmetry reversal
in the observed spectra at this frequency suggests that the higher 
frequency modes are perhaps excited higher up in the convection zone.

\section{Discussion}

We find that the observed reversal of asymmetry in the intensity power 
spectra of solar p-modes can arise as a result of contribution to the
observed flux from turbulent eddies that excite these oscillations. The
flux variations from these eddies is correlated with the p-mode oscillations
they generate (as suggested by Roxburgh \& Vorontsov 1997, Nigam et al. 1998),
causing a change to the relative contributions of the real and the 
imaginary parts of the Green's function to the power spectra and consequently
to the observed line shape. We have calculated the correlation between the
eddy flux variation and the oscillations and find it to be well within
the range required to reproduce the reversal in the line asymmetry.
The velocity power spectra of solar p-modes are unaffected by the 
contaminating noise source because the random
velocity field in the solar atmosphere, where optical line observations
are made, are completely uncorrelated with the oscillations. Intensity
power spectra on the other hand contain some contribution from
turbulent eddies at the top of the convection zone. These eddies excite 
solar p-modes as they travel downward, and the flux variation
associated with them is correlated with the oscillation they generate.

A reversal of asymmetry in the model problem considered by Roxburgh \& Vorontsov
(1997) was obtained when a superposition of dipole and quadrupole sources was 
considered. However, we find that asymmetry reversal can arise even when
modes are excited by quadrupole or dipole sources alone.
Furthermore, we have calculated the relative amplitude and 
the phase correlation in a self consistent manner, and find these
to lie in the parameter range that leads to a reversal of asymmetry. 

If the model considered here is correct then we expect the asymmetry
reversal to disappear when the observed background in the power spectra
is vanishingly small. Moreover, the use of spectral lines which
are formed deeper in the solar atmosphere, to observe solar oscillations,
should have larger amplitude for the flux variation associated with 
turbulent eddies at the top of the convection zone and hence the 
line asymmetry reversal should be more dramatic. Conversely observations
made in optical lines formed high up in the chromosphere, so that
their optical depth at the bottom of the photosphere is very large, should
show intensity and velocity power spectra to have similar shapes.
Recent work of Schou (1998) appears to support this prediction.

\acknowledgments

We are grateful to John Bahcall for encouraging us to think about
this problem, for several helpful discussions, and for his comments
on the paper. We thank J. Christensen-Dalsgaard for suggesting several
improvements to the paper and for mentioned to us the recent work of
Jesper Schou. This work was supported by a NASA grant NAG5-7395. 
SB was partially supported by an AMIAS fellowship.

\begin{figure}
\plotone{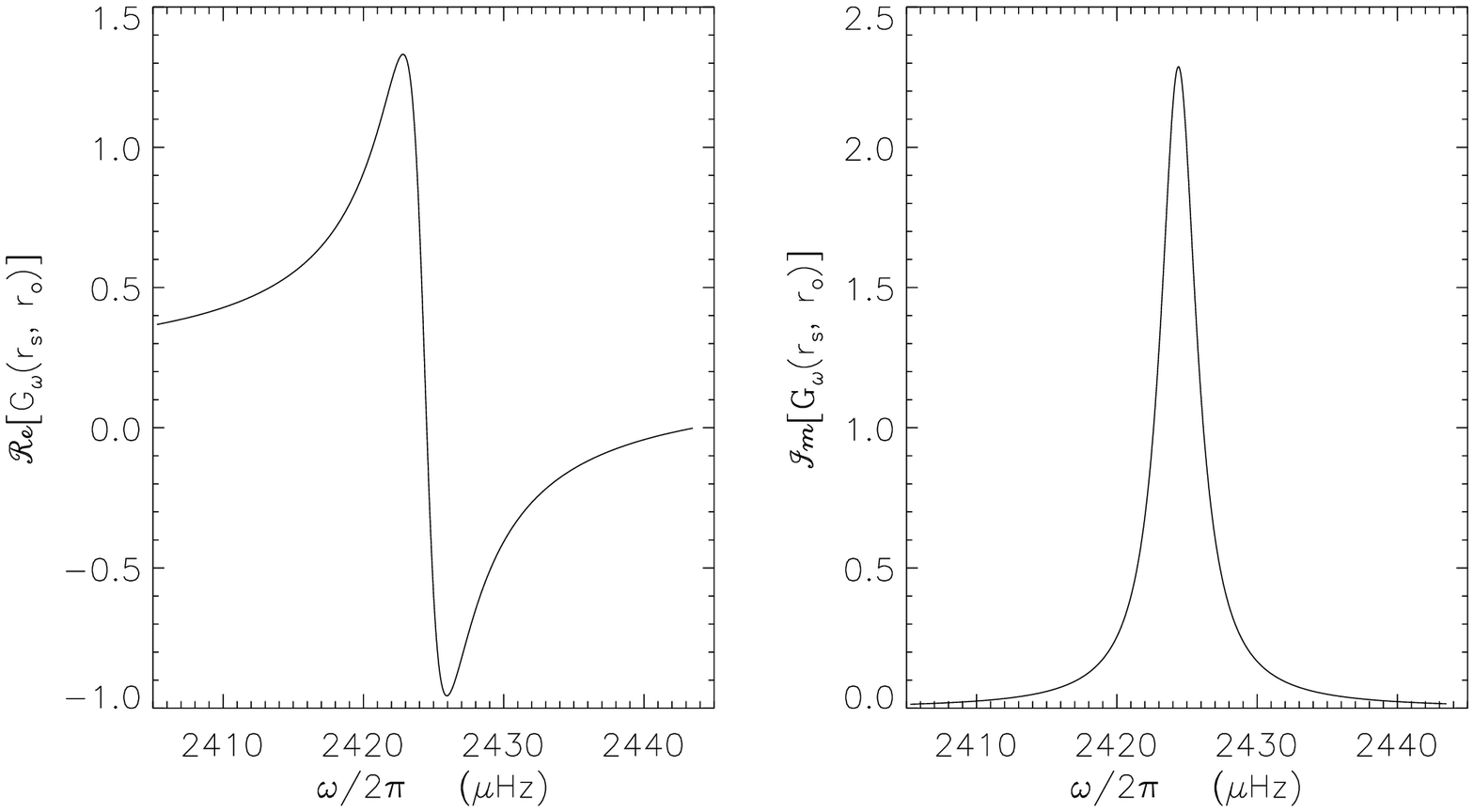}
\figcaption{ The real (left panel) and the imaginary (right panel) parts of the Green's
function, for the simple 1-D model problem, at a fixed source and observer 
location as a function of wave frequency. The source is located in the 
evanescent region approximately 200 km above the upper turning point, and
the observer is at a place corresponding to the temperature minimum in the
solar atmosphere.
}
\end{figure}

\begin{figure}
\plotone{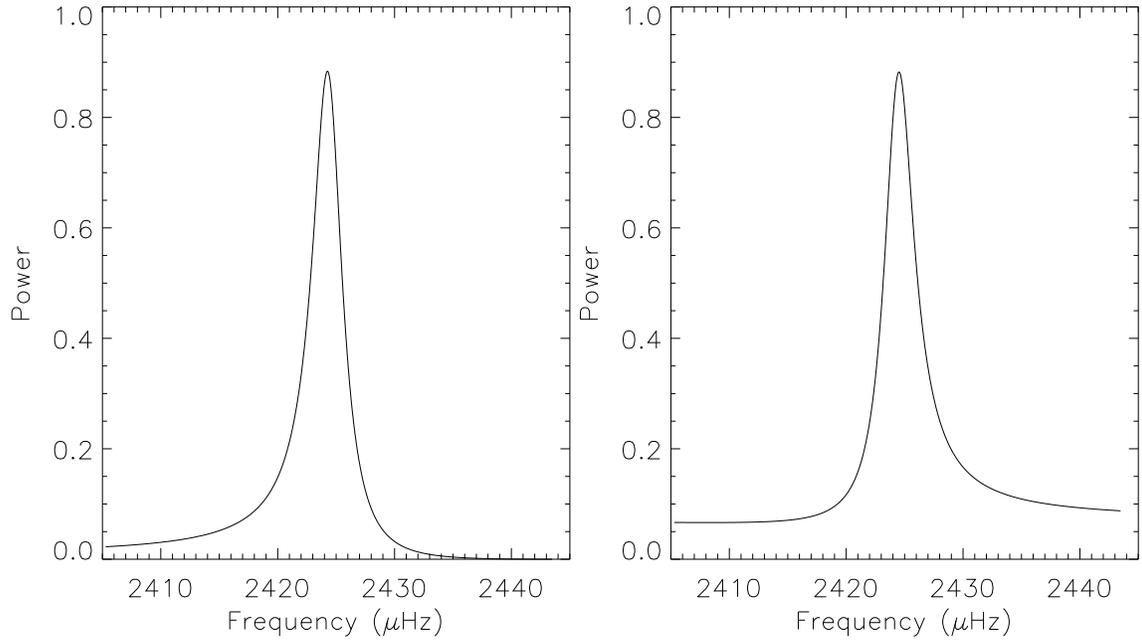}
\figcaption{
The left panel shows the velocity power spectrum for the simple
1-D model problem i.e. it does not include contribution from the noise source.
The graph in the right panel includes the flux variation from turbulent
eddies and should correspond to the intensity power spectrum; the values
of the parameters $\beta\exp(-\eta)$ and $\phi$ are 1.5 and 50$^o$ respectively.
}
\end{figure}

\begin{figure}
\plotone{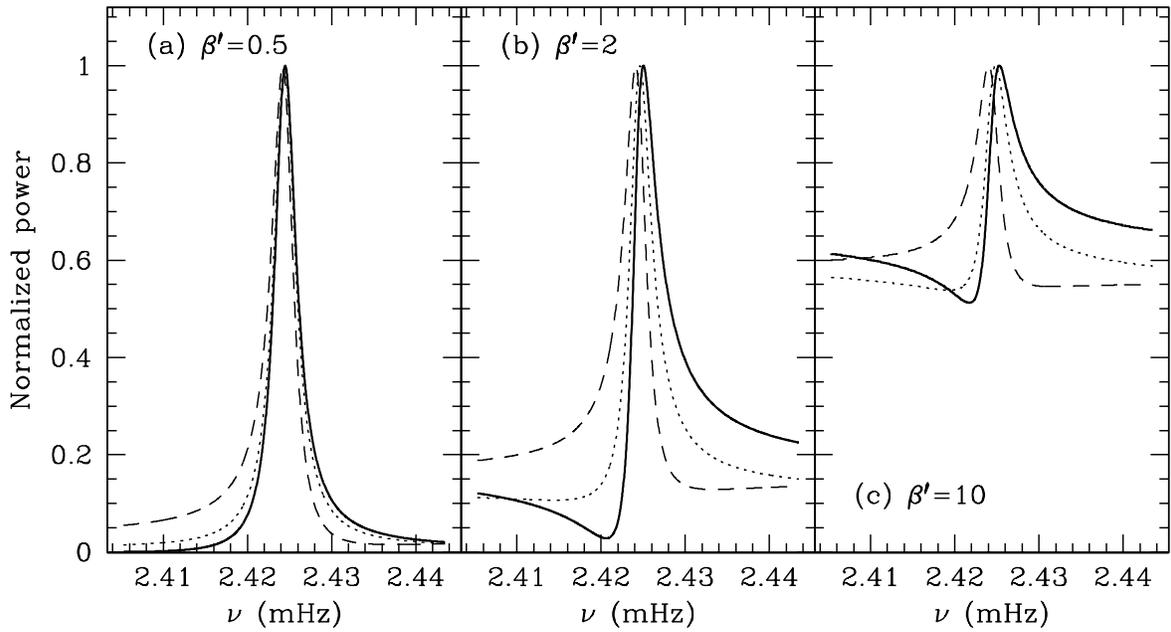}
\figcaption{
Power spectra for several different values of the parameters
($\beta, \phi$). The continuous lines are line profiles when
$\phi=0^0$, the dotted lines for $\phi=50^0$ and dashed lines
for $\phi=120^0$. For $270^o>\phi>90^o$ the sense of asymmetry
is same as in the velocity power spectrum (see fig. 2).
}
\end{figure}

\begin{figure}
\plotone{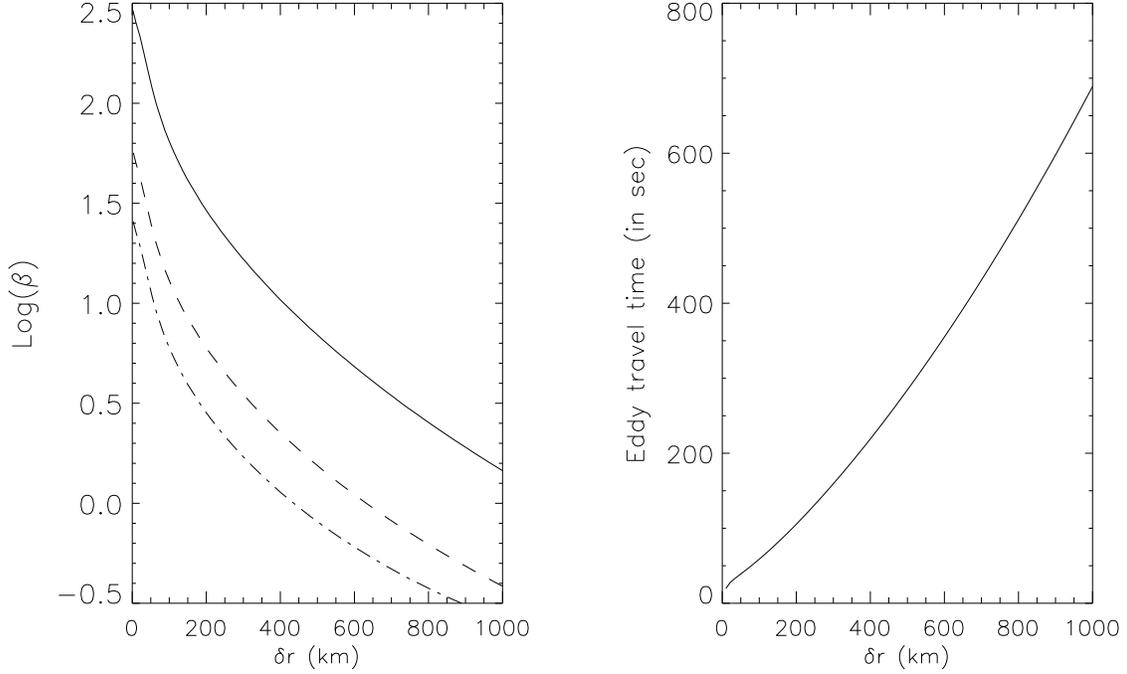}
\figcaption{
The left panel shows the value of $\beta$ for three different modes of the
sun as a function of depth measured from the top of the convection zone;
the degree ($\ell$) of all the three modes is 50, and $\nu=2.08$mHz for
the solid curve, 2.67mHz for the dashed curve and 3.2mHz for the dot-dash
curve. $\beta$ has a very weak dependence on $\ell$; it changes by 
about 20\% between $\ell$ of 2 and 50.
The panel to the right shows the eddy travel time from the
top of the convection zone ($\delta t_{tt}$) as a function of depth.
}
\end{figure}

\end{document}